\documentclass[conference]{IEEEtran}
\ifCLASSINFOpdf
\else
  \usepackage[dvips]{graphicx}
  \graphicspath{{./figures/}{./eps/}}
  \DeclareGraphicsExtensions{.eps}
\fi
\usepackage{etoolbox}
\usepackage{units}
\usepackage{microtype}
\usepackage{hyperref}
\usepackage{breakurl}
\usepackage{comment} 
\usepackage[ruled,vlined]{algorithm2e}
\DontPrintSemicolon
\SetAlFnt{\scriptsize}
\usepackage[table,xcdraw]{xcolor}
\newtoggle{inclIEEECopyRight}
\toggletrue{inclIEEECopyRight}
\newtoggle{highlightChanges}
\togglefalse{highlightChanges}
\iftoggle{highlightChanges}{
    \usepackage{changes}
}{
    \newcommand{\added}[1]{{#1}}
}
\newtoggle{doubleblind}
\togglefalse{doubleblind}
\newtoggle{includeappendix}
\togglefalse{includeappendix}
\newtoggle{includeacknowl}
\toggletrue{includeacknowl}

\newcommand{\rC}[0]{\rowcolor[HTML]{CCCCCC}}
\newcommand{\hC}[0]{\rowcolor[HTML]{333333}}
\newcommand{\tH}[1]{\multicolumn{1}{c}{\textcolor{white}{#1}}}

\makeatletter
\patchcmd{\@setref}{\bfseries ??}{\colorbox{red}{?r?}}{}{}
\patchcmd{\@citex}{\bfseries ?}{\colorbox{red}{?c?}}{}{}

\makeatother

\begin{document}
\iftoggle{inclIEEECopyRight}{
    \begin{titlepage}
    \mbox{}\\{\Large \textbf{IEEE Copyright Notice}}
    \newline\newline\newline\newline
    \textcopyright~2019 IEEE. Personal use of this material is permitted.
    Permission from IEEE must be obtained for all other uses, in any current
    or future media, including reprinting/republishing this material for
    advertising or promotional purposes, creating new collective works, for
    resale or redistribution to servers or lists, or reuse of any copyrighted
    component of this work in other works.
    \newline\newline\newline\newline
    {\large Accepted to be Published in: Proceedings of the 33rd IEEE
    International Parallel \& Distributed Processing Symposium, May 20-24,
    2019 Rio de Janeiro, Brazil}
    \end{titlepage}
}{}
\iftoggle{highlightChanges}{
    \begin{titlepage}
    \mbox{}\\{\Large \textbf{Cover Letter for Submission:}\\\\Double-precision FPUs in High-Performance Computing: an Embarrassment of Riches?}
    \newline\newline\newline\newline\large
    Summary of the implemented changes (also highlighted in blue on subsequent pages):
    \begin{itemize}
        \item Added discussion of why we report flop/s and why it is necessary in this paper
        despite our later recommendation that the HPC community should not (only) report flop/s;
        \item Split of Fig.~\ref{fig:flops} (rel./abs. flop/s comparison) into two subfigures for
        easier readability and modification of text in Sec.~\ref{ssec:eval_flops} to reflect the
        change;
        \item Added Fig.~\ref{fig:t2s-rel} for ``Time-to-Solution'' and its explanation/discussion
        in Sec.~\ref{ssec:eval_flops};
        \item Added roofline analysis in Sec.~\ref{ssec:eval_roof} to determine the optimization
        status of the FP-intensive proxy-apps, which we used for this study (incl 2 additional
        references for this part)
        \item Added details about the theoretical peak speedup with turbo boost shown in
        Fig.~\ref{fig:freq} and explanation of why a pessimistic +100Mhz was chosen in this case
        and why this resulted in ``superlinear speedup'' for some benchmarks
        \item Added acknowledgement of funding resources and author's contributions;
        \item Added note to Tab.~\ref{table:rest} to point out the multiuse of two columns by
        similar metrics (VTune reports slightly different metrics for BDW vs. KNM/KNL for
        arithmetic intensity and memory-boundedness; and readers can consult Ref. [41] for an
        in-depth documentation on these metrics (as stated previously in the table's caption));
        \item (+ multiple smaller grammar and text adjustments which will no be highlighted).
    \end{itemize}
    \end{titlepage}
}{}
\bstctlcite{IEEEexample:BSTcontrol}

%
\title{Double-precision FPUs in High-Performance Computing: an Embarrassment of Riches?}

\iftoggle{doubleblind}{
  \author{(hidden for double-blind review)}
}{
  \author{
    \IEEEauthorblockN{
  Jens Domke\IEEEauthorrefmark{1}\textsuperscript{,}\IEEEauthorrefmark{4},
  Kazuaki Matsumura\IEEEauthorrefmark{2},
  Mohamed Wahib\IEEEauthorrefmark{3},
  Haoyu Zhang\IEEEauthorrefmark{2},
  Keita Yashima\IEEEauthorrefmark{2},\\
  Toshiki Tsuchikawa\IEEEauthorrefmark{2},
  Yohei Tsuji\IEEEauthorrefmark{2},
  Artur Podobas\IEEEauthorrefmark{2}\textsuperscript{,}\IEEEauthorrefmark{4},
  Satoshi Matsuoka\IEEEauthorrefmark{4}\textsuperscript{,}\IEEEauthorrefmark{2}
}
\\
\IEEEauthorblockA{\IEEEauthorrefmark{1}Global Scientific Information and Computing Center, Tokyo Institute of Technology}
\IEEEauthorblockA{\IEEEauthorrefmark{2}Department of Mathematical and Computing Science, Tokyo Institute of Technology}
\IEEEauthorblockA{\IEEEauthorrefmark{3}AIST-TokyoTech Real World Big-Data Computation Open Innovation Laboratory, Tokyo, Japan}
\IEEEauthorblockA{\IEEEauthorrefmark{4}RIKEN Center for Computational Science (R-CCS), RIKEN, Japan}

  }
}


%


\maketitle

\begin{abstract}
Among the (uncontended) common wisdom in High-Performance Computing (HPC) is the applications' need for large amount of double-precision support in hardware.  Hardware manufacturers, the TOP500 list, and (rarely revisited) legacy software have without doubt followed and contributed to this~view. 

In this paper, we challenge that wisdom, and we do so by exhaustively comparing a large number of HPC proxy applications on two processors: Intel's Knights Landing (KNL) and Knights Mill (KNM). Although similar, the KNL and KNM architecturally deviate at one important point: the silicon area devoted to double-precision arithmetics. This fortunate discrepancy allows us to empirically quantify the performance impact in reducing the amount of hardware double-precision arithmetic. 

Our analysis shows that this common wisdom might not always be right. We find that the investigated HPC proxy applications do allow for a (significant) reduction in double-precision with little-to-no performance implications. With the advent of a failing of Moore's law, our results partially reinforce the view taken by modern industry (e.g., upcoming Fujitsu ARM64FX) to integrate hybrid-precision hardware units.

\end{abstract}


%
\IEEEpeerreviewmaketitle

\section{Introduction}\label{sec:intro}


It is becoming increasingly clear that the road forward in High-Performance Computing (HPC) is one full of obstacles.
With the ending of Dennard's scaling~\cite{dennard_design_1974} and the ending of Moore's law~\cite{moore_lithography_1995},
there is today an ever-increasing need to oversee how we allocate the silicon to various functional units in modern many-core
processors. Amongst those decisions is how we distributed the hardware support for various levels of compute-precision.

Historically, most of the compute silicon has been allocated to double-precision (DP; 64-bit) compute.
Nowadays -- in processors such as the forthcoming AA64FX~\cite{yoshida_fujitsu_2018} and NVIDIA
Volta~\cite{choquette_volta:_2018} -- the trend, mostly driven by market/AI demands, is to replace
some of the double-precision units with lower-precision units.
Lower-precision units occupy less area (up to $\approx$3x going from double- to single-precision
Fused-Multiply-Accumulate~\cite{pu_fpmax:_2016}), leading to more on-chip resources (more
instruction-level parallelism), potentially lowered energy consumption, and a definitive
decrease in external memory bandwidth pressure (i.e., more values per unit bandwidth).
The gains -- up to four times over their DP variants with little loss in
accuracy~\cite{haidar_harnessing_2018} -- are attractive and clear, but what is the impact on
performance (if any) on existing HPC applications? What performance impact can HPC users expect when migrating their code to future processors with a different distribution
in floating-point precision support? Finally, how can we empirically quantify this impact on
performance using existing processors in an apples-to-apples comparison on real-life use cases
without relying on tedious, slow, and potentially inaccurate simulators? 

The Intel Xeon Phi was supposed to be the high-end for many-core processor technology for nearly
a decade (Knights Ferry was announced in 2010), and has changed drastically since its first released.
The latest (and also last) two revisions -- the Knights Landing and Knights Mill -- are of
particular importance since they arguable reflect two different ways of thinking. Knights Landing
has relatively large support for double-precision (64-bit) computations, and follows a
more traditional school of thought. While Knights Mill follows a different direction, which is the replacement
of double-precision compute units with lower-precision (single-precision, half-precision, and integer)
compute capabilities.

In the present paper, we quantify and analyze the performance and compute bottlenecks of
Intel's Knights Landing~\cite{sodani_knights_2016} and Knights Mill~\cite{bradford_knights_2017} architectures -- two
processors with identical micro-architecture where the main difference is in the relative allocation of double-precision units.
We stress both processors with numerous realistic benchmarks from both the
Exascale Computing Project (ECP) proxy applications~\cite{noauthor_ecp_2018} and
RIKEN R-CCS Fiber Miniapp Suite~\cite{riken_aics_fiber_2015} -- benchmarks used in HPC system acquisition.
Through an extensive (and robust) performance measurement process (which we also open-source), we
empirically show the architecture's relative weaknesses. In short, the contributions of the present paper are:
\begin{enumerate}
    \item An empirical performance evaluation of the Knights Landing and Mill family of processors -- both proxies for previous and future architectural trends -- with respect to benchmarks derived from realistic HPC workloads,
    \item  An in-depth analysis of results, including identification of bottlenecks for the different application/architecture combinations, and
    \item An open-source compilation of our evaluation methodology, including our collected raw data.
\end{enumerate}

\section{Architectures, Environment, and Applications}\label{sec:materials}

Our research objective is to evaluate the impact of migrating from an architecture with (relatively) high amount of double-precision compute to an architecture with less. By high amount of double-precision compute we mean architectures whose Floating-Point Unit (FPU) has most of its silicon dedicated to 64-bit IEEE-754 floating-point operations, and by less double-precision compute we mean architectures that replace those same double-precision FPUs with lower -- potentially hybrid -- precision units.


To understand and explore the intersection of architectures with high-amount of double-precision and those with hybrid-precision, there is a need to find a processor whose architecture is unchanged with the sole exception of its floating-point unit to silicon distribution. Only one modern processor family allows for such an apples-to-apples comparison: the Xeon Phi family of processors.

\subsection{Hardware \& Software Environment}\label{ssec:hw}

\begin{table}[tbp]
    \caption{\label{table:HW} Detailed compute node hardware information; Differences between Knights Landing \& Mill highlighted in bold; Shown bandwidth (BW) measured with BabelStream (see Sec.\ref{ssec:bm}); Numbers for dual-socket reference system accumulated}
    \centering\scriptsize
    \newcommand{\tabincell}[2]{\begin{tabular}{@{}#1@{}}#2\end{tabular}}
    \begin{tabular}{|l|r|r|r|}
        \hline \hC
        \tH{Feature}                & \tH{KNL}                          & \tH{KNM}                          & \tH{Broadwell-EP} \\ \hline
        CPU Model                   & \textbf{7210F}                    & \textbf{7295}                     & 2x E5-2650v4              \\ \hline \rC
        \#\{Cores\} (HT)            & \textbf{64} (4x)                  & \textbf{72} (4x)                  & 24 (2x)                   \\ \hline
        Base Frequency              & \textbf{\unit[1.3]{GHz}}          & \textbf{\unit[1.5]{GHz}}          & \unit[2.2]{GHz}           \\ \hline \rC
        Max Turbo Freq.             & \textbf{\unit[1.5]{GHz}}          & \textbf{\unit[1.6]{GHz}}          & \unit[2.9]{GHz}           \\ \hline
        CPU Mode                    & Quadrant                          & Quadrant                          & \textit{N/A}              \\ \hline \rC
        TDP                         & \textbf{\unit[230]{W}}            & \textbf{\unit[320]{W}}            & \unit[210]{W}             \\ \hline
        DRAM Size                   & \unit[96]{GiB}                    & \unit[96]{GiB}                    & \unit[256]{GiB}           \\ \hline
        $\hookrightarrow$ Triad BW  & \textbf{\unit[71]{GB/s}}          & \textbf{\unit[88]{GB/s}}          & \unit[122]{GB/s}          \\ \hline \rC
        MCDRAM Size                 & \unit[16]{GiB}                    & \unit[16]{GiB}                    & \textit{N/A}              \\ \hline \rC
        $\hookrightarrow$ Triad BW  & \textbf{\unit[439]{GB/s}}         & \textbf{\unit[430]{GB/s}}         & \textit{N/A}              \\ \hline \rC
        MCDRAM Mode                 & Cache                             & Cache                             & \textit{N/A}              \\ \hline
        LLC Size                    & \textbf{\unit[32]{MiB}}           & \textbf{\unit[36]{MiB}}           & \unit[60]{MiB}            \\ \hline \rC
        Inst. Set Extension         & AVX-512                           & AVX-512                           & AVX2                      \\ \hline
        FP32 Peak Perf.             & \textbf{\unit[5,324]{Gflop/s}}    & \textbf{\unit[13,824]{Gflop/s}}   & \unit[1,382]{Gflop/s}     \\ \hline \rC
        FP64 Peak Perf.             & \textbf{\unit[2,662]{Gflop/s}}    & \textbf{\unit[1,728]{Gflop/s}}    & \unit[691]{Gflop/s}     \\ \hline
\end{tabular}
        \vspace{-1em}
\end{table}

Intel's Knights Landing (KNL) and Knights Mill (KNM) are the latest incarnations of a long line of architectures in the Intel's accelerator family. Both processor consist of a large number of processors cores (64 and 72, respectively), interconnected in a 2-D mesh (prior to KNL: ring interconnection). Each core has a private L1 cache and a slice of the distributed L2 cache. Caches are kept coherent through the directory-based MESIF protocol.
Both processors come with two types of external memory: MCDRAM (or, Hybrid Memory Cube) and Double-Data Rate-synchronous (DDR4) memory. Unique to the Xeon Phi processors is that the MCDRAM memory can be configured to one of three modes of operation: it is either (1) directly addressable in the global memory address space (memory-mapped), called \texttt{flat} mode, or it (2) acts as last-level cache before the DDR, called \texttt{cache} mode. Finally, the third mode (\texttt{hybrid} mode~\cite{heinecke_high_2016}) is a combination of the properties from the first two modes.

There are several policies governing where data is homed. A common high-performance configuration~\cite{gawande_scaling_2017}, which is also the one we used in our study, is the quadrant mode. Quadrant mode means that the physical cores are divided into four logical parts, where each logical part is assigned two memory controllers; each logical group is treated as a unique Non-Uniform Memory-Access (NUMA) node, allowing the operating system to perform data-locality optimizations.
Table~\ref{table:HW} surveys and contrasts the processors against each other, where the main differences are highlighted. The main architectural difference -- which is also the difference and its impact we seek to empirically quantify -- is the Floating-Point Unit (FPU). In KNL, this unit features two 512-bit wide vector units (AVX), together capable of executing 32 double-precision or 64 single-precision operations per cycle, totaling~\unit[2.6]{Tflop/s} of double- and~\unit[5.3]{Tflop/s} of single-precision peak performance, respectively, across all 64 processing cores. In KNM, however, the FPU is redesigned to replace one 512-bit vector unit with two Virtual Neural Network Instruction (VNNI) units. Those units, although specializing in hybrid-precision FMA, can execute single-precision vector instructions, but have no support for double-precision compute. Thus, in total, the KNM can execute up to~\unit[1.7]{Tflop/s} of double-precision or~\unit[13.8]{Tflop/s} of single-precision computations. In summary, the KNM has~2.59x more single-precision compute, while the KNL have~1.54x more double-precision compute.

While both the KNL and KNM are functionally and architectural similar, there are some note-worth differences.
First, the operating frequency of these processors varies: the KNL operates at a frequency of~\unit[1.3]{GHz}
(and up to~\unit[1.5]{GHz} in Turbo mode), while KNM operates at~\unit[1.5]{GHz} (\unit[1.6]{GHz} turbo).
Hence, KNM executes~15\% more cycles per second over KNL. Furthermore, although the cores of KNM and KNL are
similar (except the FPU), the number of cores is different: KNL has~64 cores while KNM has~72 cores.
Both processors are manufactured in~\unit[14]{nm} technology. Finally, the amount of on-chip last-level cache
between the two processors is different, where KNM has a~\unit[4]{MiB} advantage over KNL.

Additionally, for verification reasons, we include a modern dual-socket Xeon-based compute node in our evaluation. Despite being vastly different from the Xeon Phi systems, our Xeon Broadwell-EP (BDW) general-purpose processor is used to cross-check metrics, such as: execution time and performance (Xeon Phi should perform better), frequency-scaling experiments (BDW has more frequency domains), and performance counters (BDW exposes more performance counters).
Aside from those differences mentioned above (and highlighted in Table~\ref{table:HW}), the setup between the Xeon Phi nodes (and BDW node) is \textit{identical}, including the same operating system, software stack, and solid state disk. 

For the operating system (OS) and software environment, we use equivalent setups across our three compute nodes.
The OS is a fresh installation of CentOS~7~(minimal) with Linux kernel version 3.10.0-862, which by default has the latest versions of
the Meltdown and Spectre patches enabled. During our experiments, we limit potential OS noise by disabling all
remote storage (Network File System in our case) and allowing only a single user on the system.
Most of our applications are compiled with Intel's Parallel Studio~XE (version 2018; update 3) compilers, and we
install the latest versions of Intel TensorFlow and Intel MKL-DNN for the deep learning proxy application, since
our assumption is that Intel's software stack allows for the highest utilization of their hardware.
Exceptions to this compiler selection are listed in the subsequent Section~\ref{sec:materials}.
Furthermore, we use Intel MPI from the Parallel Studio~XE suite to execute our measurements.

\subsection{Benchmark Applications}\label{ssec:bm}




Over the years, the HPC community developed many benchmarks
that represent real workloads in order to test the capabilities of a system
-- primarily for comparisons across architectures but also for system procurement purposes.
The so-called Exascale Computing Project (ECP) proxy applications~\cite{noauthor_ecp_2018} and
RIKEN R-CCS' (f.k.a.~AICS) Fiber Miniapp Suite~\cite{riken_aics_fiber_2015}, which we will focus
on for this study, are just two examples representing modern HPC workloads.
Those benchmarks are designed to evaluate single-node and small-scale test installations,
and hence are adequate for our study.

\subsubsection{The ECP Proxy-Apps}\label{ssec:ecp}
The ECP suite (release~v1.0) consists of 12 proxy applications primarily written in C~(5x),
FORTRAN~(3x), C++~(3x), and Python~(1x), listed hereafter.

\paragraph{Algebraic multi-grid (AMG)} solver of the \textit{hypre} library is a parallel solver
for unstructured grids~\cite{park_high-performance_2015} arising from fluid dynamics problems.
We choose \textit{problem~1} for our tests, which applies a 27-point stencil on a 3-D linear system.

\paragraph{CANDLE (CNDL)} is a deep learning benchmark suite to tackle various problems in cancer
research~\cite{wozniak_candle/supervisor:_2018}.
We select benchmark 1 of pilot 1 (\textit{P1B1}), which builds an autoencoder from a sample of gene
expression data to improve the prediction of drug responses.

\paragraph{Co-designed Molecular Dynamics (CoMD)} serves as the reference implementation for
ExMatEx~\cite{mohd-yusof_co-design_2013} to facilitate co-design for (and evaluation of) classical molecular
dynamics algorithms.
We are using the included strong-scaling example to calculate the inter-atomic potential for 256,000 atoms.

\paragraph{LAGrangian High-Order Solver -- Laghos (LAGO)} computes compressible gas dynamics though
an unstructured high-order finite element method~\cite{dobrev_high-order_2012}. The input for our study is the
simulation of a 2-dimensional Sedov blast wave with default settings as documented for the
Laghos proxy-app.

\paragraph{MACSio (MxIO)} is a synthetic Multi-purpose, Application-Centric,
Scalable I/O proxy designed to closely mimic realistic I/O workloads of
HPC applications~\cite{dickson_replicating_2016}. Our input causes MACSio to write a total of \unit[433.8]{MB} to disk.

\paragraph{MiniAMR (MAMR)} is an adaptive mesh refinement proxy application of the Mantevo
project~\cite{heroux_improving_2009} which applies a stencil computation on a 3-dimensional space,
in our case a sphere moving diagonally through a cubic medium.

\paragraph{MiniFE (MiFE)} is a reference implementation of an implicit finite elements
solver~\cite{heroux_improving_2009} for scientific methods resulting in unstructured 3-dimensional grids.
For our study, we use 128$\times$128$\times$128 input dimensions for the grid.

\paragraph{MiniTri (MTri)} is able to apply different graph detection algorithms for a given graph,
such as community detection or dense subgraph detection~\cite{wolf_task-based_2015}.
As input for the triangle detection and approximation of the graph's largest clique, we download
\textit{BCSSTK30} from the MatrixMarket~\cite{boisvert_matrix_1997}.

\paragraph{Nekbone (NekB)} is a proxy for the Nek5000 application~\cite{argonne_national_laboratory_nek5000_nodate}, and uses the conjugate
gradient method for solving the standard Poisson equation for computational fluid dynamics problems.
We enabled the multi-grid preconditioner, and for strong-scaling, see Section~\ref{ssec:metrics},
we fixed the elements per process and polynomial order to one number, respectively.

\paragraph{SW4lite (SW4L)} is a proxy for the computational kernels used in the seismic modelling
software, called SW4~\cite{petersson_users_2017}, and we use the \textit{pointsource} example, which calculates the wave
propagation emitted from a single point in a half-space.

\paragraph{SWFFT (FFT)} represents the compute kernel of the HACC cosmology application~\cite{habib_hacc:_2016}
for N-body simulations. The 3-D fast Fourier transformation of SWFFT emulates
one performance-critical part of HACC's Poisson solver. In our tests, we perform 32 repetitions on a
128$\times$128$\times$128 grid.

\paragraph{XSBench (XSBn)} is the proxy for the Monte Carlo calculations used by a neutron particle transport
simulator for a Hoogenboom-Martin nuclear reactor~\cite{tramm_xsbench_2014}. We simulate a \textit{large} reactor model
represented by a \textit{unionized} grid with $15\cdot10^6$ cross-section lookups per particle.

\subsubsection{RIKEN Mini-Apps}\label{ssec:postk}
In comparison to the modernized ECP proxy-apps, RIKEN's eight mini-apps are written in
FORTRAN (4x), C (2x), and a mix of FORTRAN/C/C++ (2x).

\paragraph{FrontFlow/blue (FFB)} uses the finite element method to solve the incompressible Navier-Stokes
equation for thermo-fluid analysis~\cite{guo_basic_2006}.
We simulate the 3-D cavity flow in a rectangular space discretized into 50$\times$50$\times$50 cubes.

\paragraph{Frontflow/violet Cartesian (FFVC)} falls into the same problem class as
FFB, however the difference is that FFVC uses the finite volume method (FVM)~\cite{ono_ffv-c_nodate}.
Here, we calculate the 3-D cavity flow in a 144$\times$144$\times$144 cuboid.

\paragraph{MODYLAS (MDYL)} makes use of the fast multipole method for long-range force evaluations in
molecular dynamics simulations~\cite{andoh_modylas:_2013}.
Our input is the \textit{wat222} example which distributes 156,240 atoms over a
16$\times$16$\times$16 cell domain.

\paragraph{many-variable Variational Monte Carlo (mVMC) method} implemented by this mini-app is used
to simulate quantum lattice models for studying the physics of condensed matter~\cite{misawa_mvmc--open-source_2018}.
We use mVMC's included strong-scaling test, but downsize it (1/3 lattice dimensions and 1/4 of samples). 

\paragraph{Nonhydrostatic ICosahedral Atmospheric Model (NICM)} is a proxy of NICAM, which
computes mesoscale convective cloud systems based on FVM for icosahedral grids~\cite{tomita_new_2004}.
We run Jablonowski's baroclinic wave test (\textit{gl05rl00z40pe10}), but reduce the
simulated days from~11~to~1.

\paragraph{Next-Gen Sequencing Analyzer (NGSA)} is a mini-app of a genome analyzer and a set of
alignment tools designed to facilitate cancer research by detecting genetic mutations in
human DNA~\cite{riken_csrp_grand_2013}.
For our experiments, we rely on pre-generated pseudo-genome data (\textit{ngsa-dummy}).

\paragraph{NTChem (NTCh)} implements a computational kernel of the NTChem software framework
for quantum chemistry calculations of molecular electronic structures, i.e., the solver for the
second-order M{\o}ller-Plesset perturbation theory~\cite{nakajima_ntchem:_2014}. We select the
H\textsubscript{2}O test case for our study.

\paragraph{Quantum ChromoDynamics (QCD)} mini-app solves the lattice QCD problem in a 4-D
lattice (3-D plus time), represented by a sparse coefficient matrix, to investigate the
interaction between quarks~\cite{boku_multi-block/multi-core_2012}. We evaluate QCD with the \textit{Class 2}
input for a $32^3 \times 32$ lattice discretization.

\begin{table}[tbp]
    \caption{\label{table:APP} Application Categorization, Compute Patterns, and main Programming Languages used; MACSio, HPL, HPCG, and BabelStream Benchmarks omitted}
    \centering\scriptsize
    \begin{tabular}{|l|l|l|l|}
        \hline \hC
        \tH{ECP}    & \tH{Scientific/Engineering Domain}    & \tH{Compute Pattern}  & \tH{Language} \\ \hline
        AMG         & Physics and Bioscience                & Stencil               & C \\ \hline \rC
        CANDLE      & Bioscience                            & Dense matrix          & Python \\ \hline
        CoMD        & Material Science/Engineering          & N-body                & C \\ \hline  \rC
        Laghos      & Physics                               & Irregular             & C++\\ \hline
        miniAMR     & Geoscience/Earthscience               & Stencil               & C \\ \hline \rC
        miniFE      & Physics                               & Irregular             & C++ \\ \hline
        miniTRI     & Math/Computer Science                 & Irregular             & C++ \\ \hline \rC
        Nekbone     & Math/Computer Science                 & Sparse matrix         & Fortan \\ \hline
        SW4lite     & Geoscience/Earthscience               & Stencil               & C \\ \hline \rC
        SWFFT       & Physics                               & FFT                   & C/Fortran \\ \hline
        XSBench     & Physics                               & Irregular             & C \\ \hline\hline \hC
        \tH{RIKEN}  & \tH{Scientific/Engineering Domain}    & \tH{Compute Pattern}  & \tH{Language} \\ \hline
        FFB         & Engineering (Mechanics, CFD)          & Stencil               & Fortran \\ \hline \rC
        FFVC        & Engineering (Mechanics, CFD)          & Stencil               & C++/Fortran \\ \hline
        mVMC        & Physics                               & Dense matrix          & C \\ \hline \rC
        NICAM       & Geoscience/Earthscience               & Stencil               & Fortran \\ \hline
        NGSA        & Bioscience                            & Irregular             & C \\ \hline \rC
        MODYLAS     & Physics and Chemistry                 & N-body                & Fortran \\ \hline
        NTChem      & Chemistry                             & Dense matrix          & Fortran \\ \hline \rC
        QCD         & Lattice QCD                           & Stencil               & Fortran/C \\ \hline
    \end{tabular}
    \vspace{-1em}
\end{table}

\subsubsection{Reference Benchmarks}\label{ssec:refbm}

In addition to those 20 applications, we use the compute intensive HPL~\cite{dongarra_linpack_1988} benchmark,
and HPCG~\cite{dongarra_new_2016} and stream (both memory intensive) to evaluate
the baseline of the investigated architectures. 

\paragraph{High Performance Linpack (HPL)} is solving a dense system of linear equations $Ax = b$
to demonstrate the double-precision compute capabilities of a (HPC)
system~\cite{strohmaier_top500_2018}. Our problem size is 64,512.
For both HPL and HPCG (see below), we employ highly tuned versions shipped with Intel's Parallel Studio
XE suite with appropriate parameters for our systems.

\paragraph{High Performance Conjugate Gradients (HPCG)} is applying a conjugate gradient solver
to a system of linear equation (sparse matrix $A$), with the intent to
demonstrate the system's memory subsystem and network limits. We choose 360$\times$360$\times$360 as
global problem dimensions for HPCG.

\paragraph{BabelStream (BABL)} is one of many available ``stream'' benchmarks supporting
evaluations of the memory subsystem for CPUs and accelerators~\cite{deakin_gpu-stream_2016}.
We will use~\unit[2]{GiB} and~\unit[14]{GiB} input vectors, see Section~\ref{ssec:eval_mem}
for details.

We provide a compressed overview of the ECP and RIKEN's proxy applications in Table~\ref{table:APP}.
In this table, each application is categorized by its scientific domain, as well as the primary
workload/kernel classification, for which we use the classifiers employed by Hashimoto et al.~\cite{hashimoto_empirical_2017}.
Both, the scientific domain as well as the kernel classification will be important for our subsequent
analysis in Sections~\ref{sec:eval} and~\ref{sec:discuss}.

\section{Methodology}\label{sec:methods}
%
In this section, we present our rigorous benchmarking approach into investigating the characteristics of each architecture, and extracting the necessary information for our study.
%

\subsection{Benchmark Setup and Configuration Selection}\label{ssec:bmconf}

Due to the fact that the benchmarks, listed in Section~\ref{ssec:bm}, are firstly realistic proxies of the
original applications~\cite{aaziz_methodology_2018} and secondly are used in the procurement process, we can assume
that these benchmarks are well tuned and come with appropriate compiler options for a variety of compilers -- \added{a hypothesis we will test in Section~\ref{ssec:eval_roof}}.
Hence, we refrain from both manual code optimization and alterations of the compiler options.
%
%
The only modifications we perform are:
\begin{itemize}
    \item Enabling interprocedural optimization (\texttt{-ipo}) and compilation for the highest instruction set available (\texttt{-xHost})\footnote{~Exceptions: (a) AMG compiled with \texttt{-xCORE-AVX2} to avoid arithmetic \\$~~~\,\quad$errors; (b) NGSA's BWA tool compiled with GNU gcc to avoid segfaults.},
    \item Patching a segmentation fault in MACSio\footnote{~After our reporting, the developers patched the upstream version.}, and
    \item Injecting our measurement source code, see Section~\ref{ssec:metrics}.
\end{itemize}
With respect to the measurement runs, we follow this five step approach for each benchmark:
\begin{enumerate}
    \item[0)] Install, patch, and compile the benchmark, see above, 
    \item Select appropriate inputs/parameters/seeds for execution,
    \item Determine ``best'' parallelism: \#processes and \#threads,
    \item Execute a \textit{performance}, a \textit{profiling}, and a \textit{frequency} run,
    \item Analyze the results (go to 0. if anomalies are detected).
\end{enumerate}
and we will further elaborate on those steps hereafter.

%
%
For the input selection we have to balance between multiple constraints and choose based on: Which
recommended inputs are listed by the benchmark developers?, How long does the benchmark run?\footnote{~Our
aim is~\unit[1]{sec}--\unit[10]{min} due to the large sample size we have to cover.} Does it occupy a
realistic amount of main memory (e.g., avoid cache-only executions)? Are the results repeatable
(randomness/seeds)? We optimize for the metrics reported by the benchmark (e.g., select the input
with the highest~\unit[]{Gflop/s} rate).

%
Furthermore, one of the most important consideration while selecting the right inputs is
\textit{strong-scaling}. We require strong-scaling properties of the benchmark for two reasons:
the results collected in Step~(2) need to be comparable, and even more importantly, the results
of Step (3) must be comparable between different architectures, since we may have to use different
numbers of MPI processes for KNL and KNL (and our BDW reference architecture) due to their difference
in core counts. The only exception is MiniAMR for which we are unable to find a strong-scaling
input configuration and instead optimized for the reported~\unit[]{Gflop/s} of the benchmark.
Accordingly, we then choose the same amount of MPI processes on our KNL and KNM compute nodes for MiniAMR.

In Step (2), we evaluate numerous combinations of MPI processes and OpenMP threads
for each benchmark, including combinations which over-/undersubscribe the CPU cores, and test each
combination with three runs to minimize the potential for outliers due to system noise.
For all subsequent measurements, we select the number of processes and threads based on the ``best'' (w.r.t
time-to-solution of the solver) combination among these tested versions, see Table~\ref{table:rest} 
at the end of this paper for details.
We are not applying specific tuning options to Intel's MPI library, except for using Intel's recommended
settings for HPCG with respect to thread affinity and MPI\_allreduce. 
The reason is that our pretests (with a subset of the benchmarks) with non-default parameters for
Intel MPI consistently resulted in longer time-to-solution.

%
For Step (3), we run each benchmark ten times to identify the fastest time-to-solution for the
(compute) kernel of the benchmark. Additionally, for the profiling runs, we execute the benchmark
once for each of the profiling tools and/or metrics (in case the tool is used for multiple metrics),
see Section~\ref{ssec:metrics} for details. Finally, we perform frequency scaling experiments
for each benchmark, where we throttle the CPU frequency to all of the available lower CPU states
below the maximum CPU frequency, which we use for the performance runs, and record the lowest kernel
time-to-solution among ten trials per frequency. The reason and results of the frequency scaling
test will be further explained in Section~\ref{ssec:eval_freq}.
One may argue for more than ten runs per benchmark to find the optimal time-to-solution, however,
given the prediction interval theory and our deterministic benchmarks executed on a single node,
it is unlikely to obtain a much faster run and we confirmed that the fastest~50\% of executions per
benchmark only vary by~3.9\% on average.
The collected metrics, see the following section, will be analyzed in Section~\ref{sec:eval} in detail.

\subsection{Metrics and Measurement Tools}\label{ssec:metrics}
%
%
%
To study and analyze the floating point requirements by applications, it is not only important to
evaluate an established metric (floating point operations per second), but also other metrics,
such as memory throughput, cache utilization, or speedup with increased CPU frequency.
The detailed list of metrics (and derived metrics) and the methodology and tools we use to
collect these metrics will be explained hereafter.

One observation is that the amount of time spent on initializing and post processing within
each proxy application can be relatively high (e.g., HPCG spends only~11\% and~30\% of its
time in the solver part on BDW and Phi, respectively) and is usually not consistent with the real workloads, e.g., one can
reduce the epochs for performance evaluation purposes in CANDLE but not the input data
pre-processing to execute those epochs. These mismatches in kernel-to-[pre$|$post]processing ratio
requires us to extract all metrics only for the (computational) kernel of the benchmark.
Hence, we identify and inject profiling instructions around the kernels to start or pause the
collection of raw metric data by the analysis tools. This code injection is exemplified in
PseudoCode~\ref{code:inj}. Therefore, unless otherwise stated in this Section or subsequent sections,
all presented data will be based exclusively on the kernel portion of each benchmark.
\begin{algorithm}[tbp]
    \label{code:inj}
    \SetKwProg{Fn}{Function}{ is}{end}
    \#define START\_ASSAY \{measure time; toggle on [PCM $|$ SDE $|$ VTune]\} \;
    \#define STOP\_ASSAY \{measure time; toggle off [PCM $|$ SDE $|$ VTune]\} \;
    \Fn{main}{
        STOP\_ASSAY\;
        {\color{gray}Initialize benchmark}\;
        \ForEach{{\color{gray} solver loop}}{
            START\_ASSAY\;
            {\color{gray}Call benchmark solver/kernel}\;
            STOP\_ASSAY\;
            {\color{gray}Post-processing}\;
        }
        {\color{gray}Verify benchmark result}\;
        START\_ASSAY\;
    }
    \caption{Injecting analysis instructions}
    \vspace{-.5em}
\end{algorithm}

%
For tool stability reason, attention to detail/accuracy, and overlap with our needs, we settle
on the use of the MPI API for runtime measurements, alongside with Intel's Processor Counter
Monitor (PCM)~\cite{willhalm_intel_2017}, Intel's Software Development Emulator (SDE)~\cite{raman_calculating_2015}, and Intel's VTune
Amplifier~\cite{sobhee_intel_2018}\footnote{~To avoid persistent compute node crashes (likely due to incompatibilities\\$~~~\,\quad$with the Spectre/Meltdown patches), we had to disable VTune's build-in 
\\$~~~\,\quad$sampling driver and instead rely on Linux' \texttt{perf} tool.}.
Furthermore, as auxiliary tools we rely on RRZE's Likwid~\cite{treibig_likwid:_2010} for frequency
scaling\footnote{~Our Linux kernel version required us to disable
the default Intel P-State\\$~~~\,\quad$driver to have full access to the fine-grained frequency scaling.} and
LLNL's msr-safe~\cite{walker_best_2016} for allowing access to CPU model-specific registers.
An overview of (raw) metrics which we extract with these tools for the benchmarks, listed in
Section~\ref{ssec:bm}, is shown in Table~\ref{tb:Mtools}. Furthermore, derived metrics, such
as~\unit[]{Gflop/s}, will be explained on-demand in Section~\ref{sec:eval}.
\begin{table}[tp]
    \centering\scriptsize
    \caption{\label{tb:Mtools}Summary of metrics and method/tool to collect these metrics}
    \begin{tabular}{|l|l|}
        \hline \hC
        \multicolumn{1}{c}{\textcolor{white}{Raw Metric}}   & \multicolumn{1}{c}{\textcolor{white}{Method/Tools}}   \\\hline
        Runtime [\unit[]{s}]                & MPI\_Wtime()                      \\\hline \rC
        \#\{FP / integer operations\}       & SDE                               \\\hline
        \#\{Branches operations\}           & SDE                               \\\hline \rC
        Memory throughput [\unit[]{B/s}]    & PCM (pcm-memory.x)                \\\hline
        \#\{L2/LLC cache hits/misses\}      & PCM (pcm.x)                       \\\hline \rC
        Consumed Power [\unit[]{Watt}]      & PCM (pcm-power.x)                 \\\hline
        SIMD instructions per cycle         & perf + VTune (`hpc-performance')  \\\hline \rC
        Memory/Back-end boundedness         & perf + VTune (`memory-access')    \\\hline
    \end{tabular}
    \vspace{-.5em}
\end{table}

\section{Evaluation}\label{sec:eval}
%
The following subsections will primarily focus on visualizing and analyzing the key metrics we collect for each proxy- and mini-app,
such as \unit[]{Gflop/s}. The significance of our findings with respect to future software, CPU, and HPC system design will then
be discussed in the next Section~\ref{sec:discuss}.
\added{
While we will argue for less \unit[]{flop/s}-centric performance reporting of
HPC benchmarks in Section~\ref{sec:discuss}, we have to adhere the current standards.
By analyzing the performed FP operation/s instead of concealing them, not only do we gain insight
into realistic \unit[]{flop/s} of HPC applications, we also have the capability to evaluate FP unit requirements.
Furthermore, this analysis will strengthen our argument that \unit[]{flop/s} should not be the only
reported performance metric -- especially if the majority of benchmarks does not even achieve 10\% of theoretical peak.
}

Furthermore, analyzing other metrics such as the instruction mix, time-to-solution, or memory throughput,
see Section~\ref{ssec:eval_ops},~\ref{ssec:eval_flops}, and~\ref{ssec:eval_mem},
in an isolated fashion does also not give good indications about the system's bottlenecks,
and hence, especially when reasoning about FPU requirements, we have to understand
the applications' compute-boundedness, which we evaluate in Section~\ref{ssec:eval_freq}.
Only when analyzing all these metrics in the same context, we attain the needed understanding.
Table~\ref{tb:Mtools} summarizes the primary metrics and method/tool used to collect these metrics,
while Table~\ref{table:rest} includes additionally collected metrics.

\subsection{Integer vs. Single-Precision FP vs. Double-Precision FP}\label{ssec:eval_ops}
\begin{figure}[tbp]
    \centering
    \includegraphics[width=\linewidth]{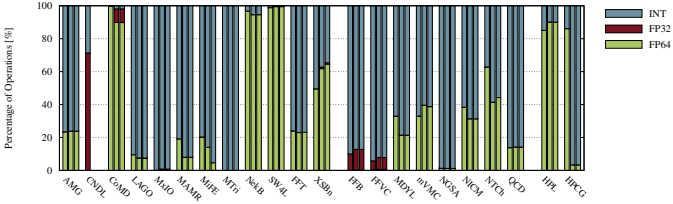}
    \caption{\label{fig:totalops} Ratio of integer vs. single-precision FP vs. double-precision FP per proxy-app as counted by Intel's SDE; Per application: \textbf{Left bar = BDW, middle bar = KNL, right bar = KNM}; Missing bars for CANDLE due to SDE crashes on Xeon Phi; Proxy-app abbreviations acc. to Section~\ref{ssec:bm}}
    \vspace{-0.6em}
\end{figure}
%
The breakdown of total number of integer and single/double-precision floating point (FP) operations, as depicted in Figure~\ref{fig:totalops},
shows two rather unexpected trends. First, the number of proxy-apps relying on 32-bit FP instructions is four out of 22, which is surprisingly low, and furthermore, only one of them utilizes both 32-bit and 64-bit FP instructions.
Minor variances in integer to FP ratio between the architectures can likely be explained by the difference in
AVX vector length, quality of compiler optimization for each CPU, and execution/parallelization approach.
The second unexpected trend is the imbalance of integer to FP operations, i.e., 16 of 22 applications issue at least 50\%
integer operations. However, one has to keep in mind that the Intel SDE output includes AVX vector instructions for integers, where
the granularity can be as low as 1-bit per operand (cf. 4 or \unit[8]{byte} per FP operand). Hence, the total integer operations
count might be slightly inflated.
Lastly, the results for HPCG show a big discrepancy between BDW and KNL/KNM. While the total FP operations count is similar,
Intel's optimized binary for KNL/KNM issues far more integer operations, see Table~\ref{table:rest} for details, and we are unaware of the reason.

\subsection{Floating-Point Operation/s and Time-to-Solution}\label{ssec:eval_flops}
\begin{figure}[tbp]
    \centering
    \includegraphics[width=\linewidth]{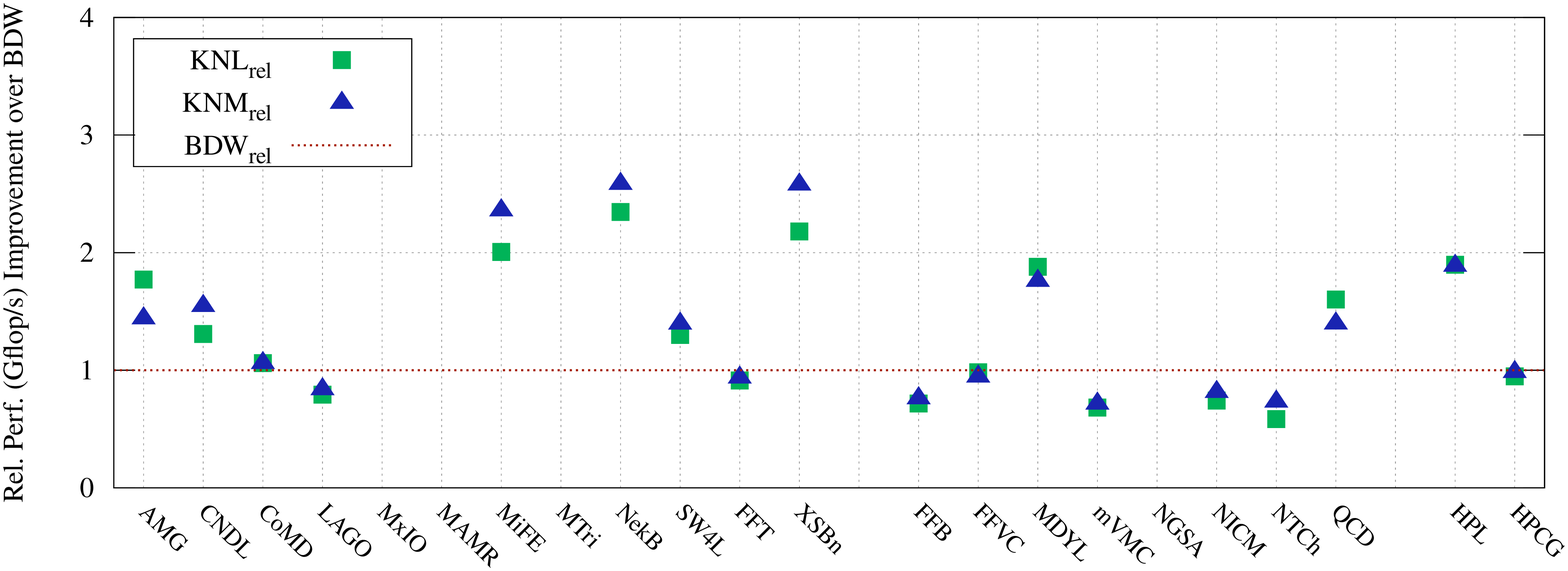}
    \includegraphics[width=\linewidth]{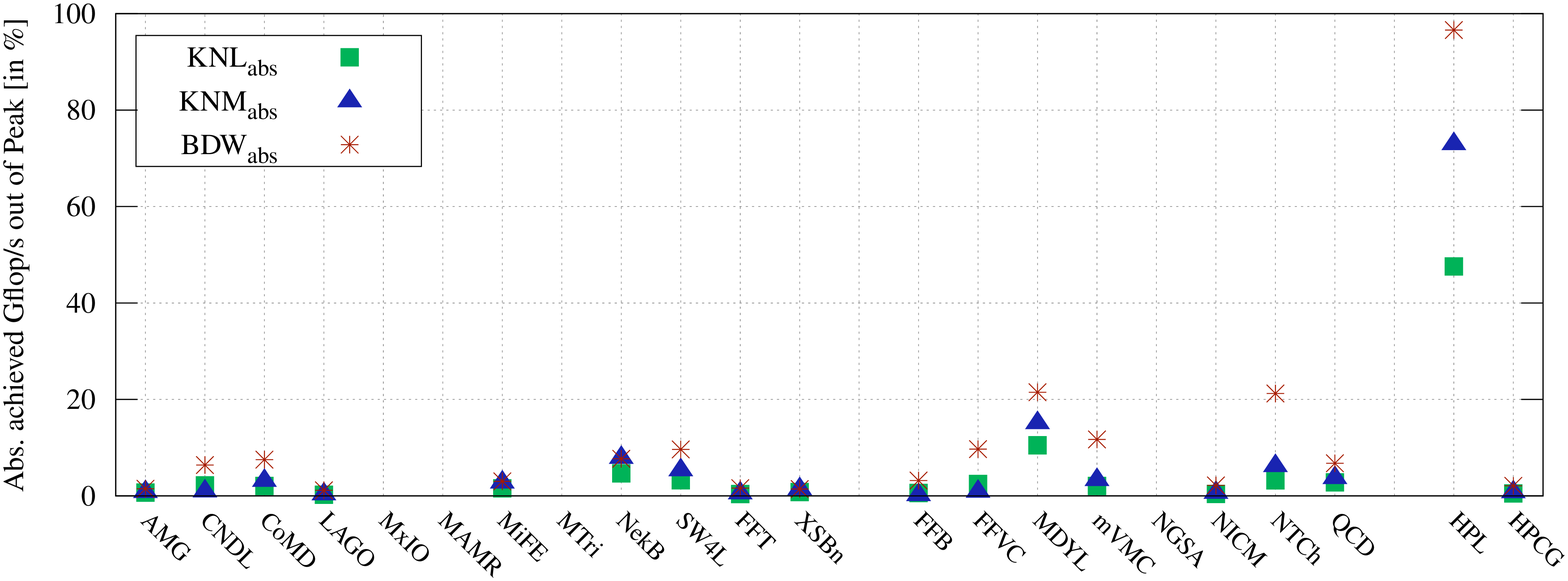}
    \caption{\label{fig:flops} \added{\textbf{Relative floating-point performance} (FP32 and FP64 \unit[]{Gflop/s} accumulated) of KNL and KNM in comparison to dual-socket Broadwell-EP (see \textbf{top plot}) and \textbf{Absolute achieved \unit[]{Gflop/s}} w.r.t dominant FP operations (cf. Fig.~\ref{fig:totalops}) in comparison to theoretical peak performance listed in Tab.~\ref{table:HW} (see \textbf{bottom plot})}; Due to missing SDE data for CANDLE, we assume the total number of FP operations is equivalent to BDW and divide by CANDLE's time-to-solution; Filtered proxy-apps with negligible FP operations: MxIO, MTri, and NGSA; Filtered out MiniAMR because of the strong-scaling issue described in Section~\ref{ssec:bmconf}; Proxy-app abbreviations acc.~to Section~\ref{ssec:bm}}
    \vspace{-0.3em}
\end{figure}
\begin{figure}[tbp]
    \centering
    \includegraphics[width=\linewidth]{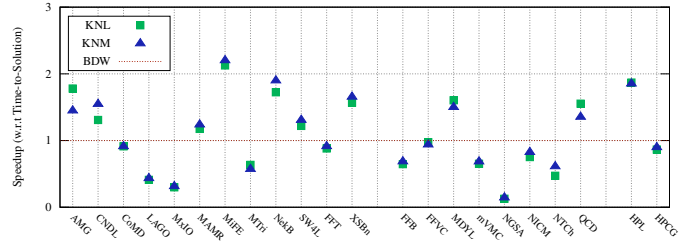}
    \caption{\label{fig:t2s-rel} \added{Runtime speedup of KNL/KNM in comparison to dual-socket Broadwell-EP; MiniAMR included, but only KNL-to-KNM comparison valid due to differences in \#\{MPI processes\} and aforementioned strong-scaling issues (see Section~\ref{ssec:bmconf}); Proxy-app abbreviations acc.~to Section~\ref{ssec:bm}}}
    \vspace{-0.9em}
\end{figure}
Figure~\ref{fig:flops} shows the relative performance improvement of KNL/KNM over the dual-socket BDW node and the absolute achieved~\unit[]{Gflop/s} on each processor. It is important to note that all proxy-/mini-apps, with the exception of HPL, have less than~21.5\% (BDW), 10.5\% (KNL), and~15.1\%~(KNM) FP efficiency. Given that these applications are presumably optimized, and still achieve this low FP efficiency, implies a limited relevance of FP unit's availability. The figure shows that the majority of codes have comparable performance on KNM versus KNL. Notable mentions are: a)~CANDLE which benefits from VNNI units in mixed precision,
b)~MiFE, NekB, and XSBn which improve probably due to increased core count and KNM's higher CPU frequency,
and c)~some memory-bound applications (i.e., AMG, HPCG, and MTri) which get slower supposedly due to the difference in peak throughput demonstrated in Figure~\ref{fig:memthru} in addition to the increased core count causing higher competition for bandwidth.

\added{
While we filtered out applications which do not perform a significant amount of FP operations in
Figure~\ref{fig:flops}, we added these applications in the time-to-solution comparison, shown in
Figure~\ref{fig:t2s-rel}, to gain a more comprehensive view.
Overall, the speedups of KNL/KNM over our reference system match the expectation we
reached from Figure~\ref{fig:flops} (top plot). However, one noticeable outlier is Laghos,
which is caused by the application executing $\approx$2x more FP64 operations on KNL/KNM,
but also running about two times longer, and hence \unit[]{flop/s} are roughly the same,
while the time-to-solution differs from BDW.
}

\subsection{Memory Throughput of (MC-)DRAM}\label{ssec:eval_mem}
\begin{figure}[tbp]
    \centering
    \includegraphics[width=\linewidth]{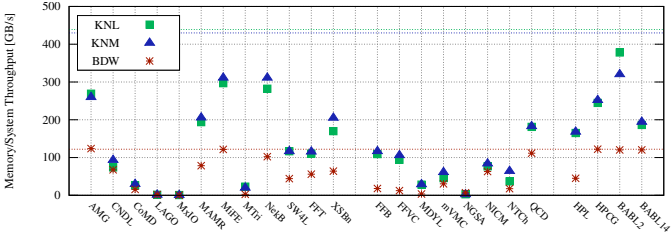}
    \caption{\label{fig:memthru} Memory throughput (only DRAM for BDW, DRAM+MCDRAM for Phi) per proxy-app; Dotted lines indicate Triad stream bandwidth (flat mode, cf. Tab.~\ref{table:HW}); BabelStream for \unit[2]{GiB} (BABL2) and \unit[14]{GiB} (BABL14) vector length added (measured in cache mode); Proxy-app labels acc.~to Section~\ref{ssec:bm}}
    \vspace{-0.2em}
\end{figure}
For the memory throughput measurements, shown in Figure~\ref{fig:memthru}, we use Intel's PCM tool to analyze DRAM and MCDRAM throughput.
Our measurements with BabelStream are included as well to demonstrate the maximum achievable bandwidth, see horizontal lines for MCDRAM
(in \textit{flat} mode), which is lower when the MCDRAM is used in \textit{cache} mode.
We still achieve~86\% on KNL and~75\% on KNM when the vectors fit into MCDRAM, but drop to slightly
higher than DRAM throughput (due to minor prefetching benefits) when the vectors do not fit (see BABL14 for \unit[14]{GiB} vectors).
This throughput advantage of the MCDRAM translates into a performance boost for six proxy-apps (AMG, MAMR, MiFE, NekB, XSBn, and QCD; in comparison to BDW)
which heavily utilize the available bandwidth, see Figure~\ref{fig:memthru}, and which are memory-bound on our reference system.
This can easily be verified when comparing the time-to-solution for these kernels as show in Figure~\ref{fig:t2s-rel} and broken down into numbers in Table~\ref{table:rest}.
Only HPCG cannot benefit from the higher bandwidth and, despite showing $\approx$2x throughput, the runtime drops by more than 10\%,
indicating a memory-latency issue of HPCG on KNL/KNM, which is one of the design goals for the benchmark~\cite{dongarra_new_2016}.

\subsection{Roofline Model Analysis for Broadwell-EP System}\label{ssec:eval_roof}
\begin{figure}[tbp]
    \centering
    \includegraphics[width=\linewidth]{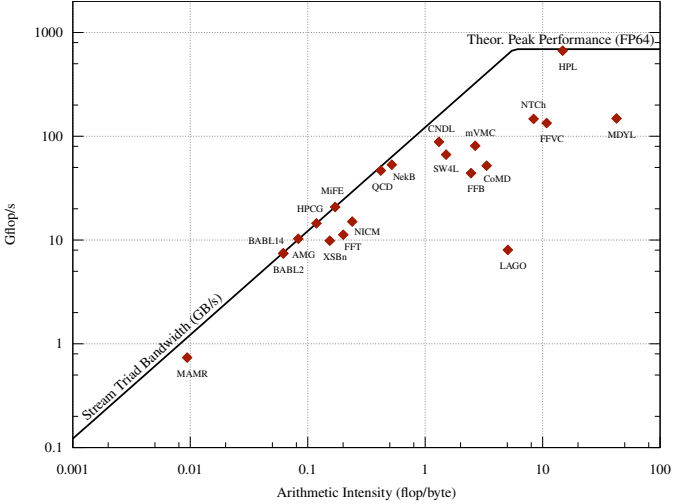}
    \caption{\label{fig:roof} \added{Roofline plot (w.r.t dominant FP operations and DRAM bandwidth) for Broadwell-EP reference system; Filtered proxy-apps with negligible FP operations: MxIO, MTri, and NGSA; Proxy-app labels acc.~to Section~\ref{ssec:bm}}}
    \vspace{-0.8em}
\end{figure}
\added{
Based on the collected data of Sections~\ref{ssec:eval_flops} and~\ref{ssec:eval_mem}, we calculate 
the location of each of our FP-intensive proxies (including BabelStream Triad) within the roofline
graph for our x86-based reference system, as visualized in Figure~\ref{fig:roof}.
These results match largely our expectations about the applications' optimizations for x86-based 
architectures, as well as similar analysis performed for other HPC workloads and
benchmarks~\cite{asifuzzaman_report_2017,jouppi_-datacenter_2017,ofenbeck_applying_2014}.
The only noticeable outlier is Laghos, which leaves room for performance tuning, and hence 
challenges our initial assumption of Section~\ref{ssec:bmconf}.
}

\added{
Given that almost all proxy-apps are in fact optimized for x86, and that both KNL and KNM are x86 ISA,
likely all levels of (threads-, vector-, instruction-) parallelism are already exposed.
The remaining question is how well the runtime and the compiler utilizes this parallelism.
We covered both aspects by our approach (see Section~\ref{ssec:bmconf}) of determining
the best combination of MPI processes and OpenMP threads and instructing the compiler to 
optimize for the host architecture. Consequently, our roofline plots for KNL/KNM reveal
similar information, and are therefore omitted.
}

\subsection{Frequency Scaling to Identify Compute-Boundedness}\label{ssec:eval_freq}
\begin{figure}[tbp]
    \centering
    \includegraphics[width=\linewidth]{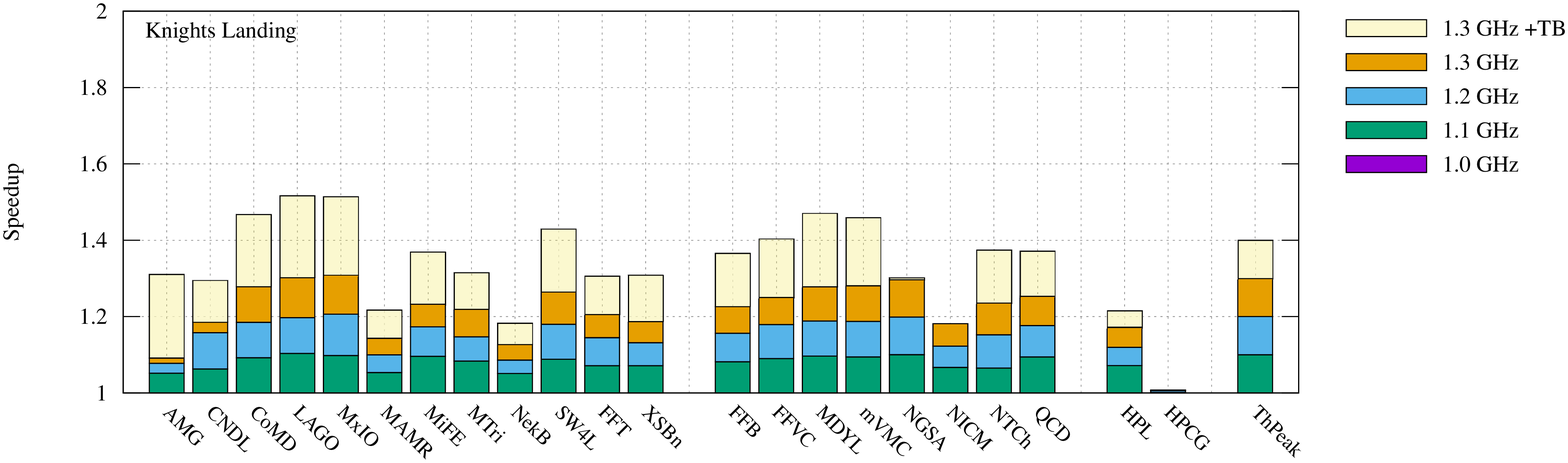}
    \includegraphics[width=\linewidth]{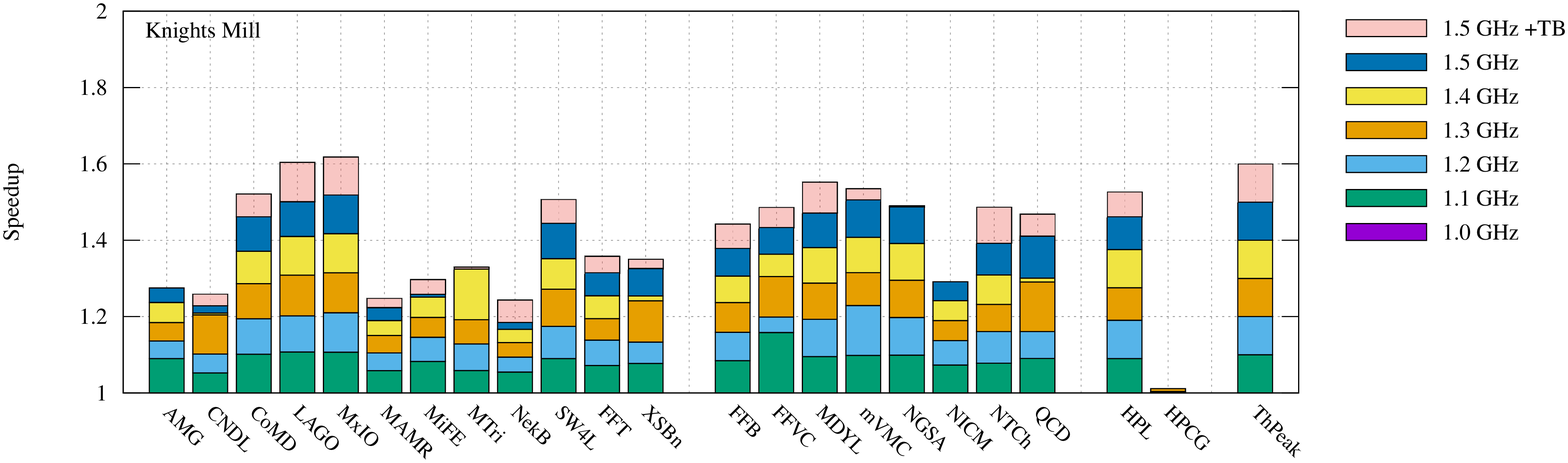}
    \includegraphics[width=\linewidth]{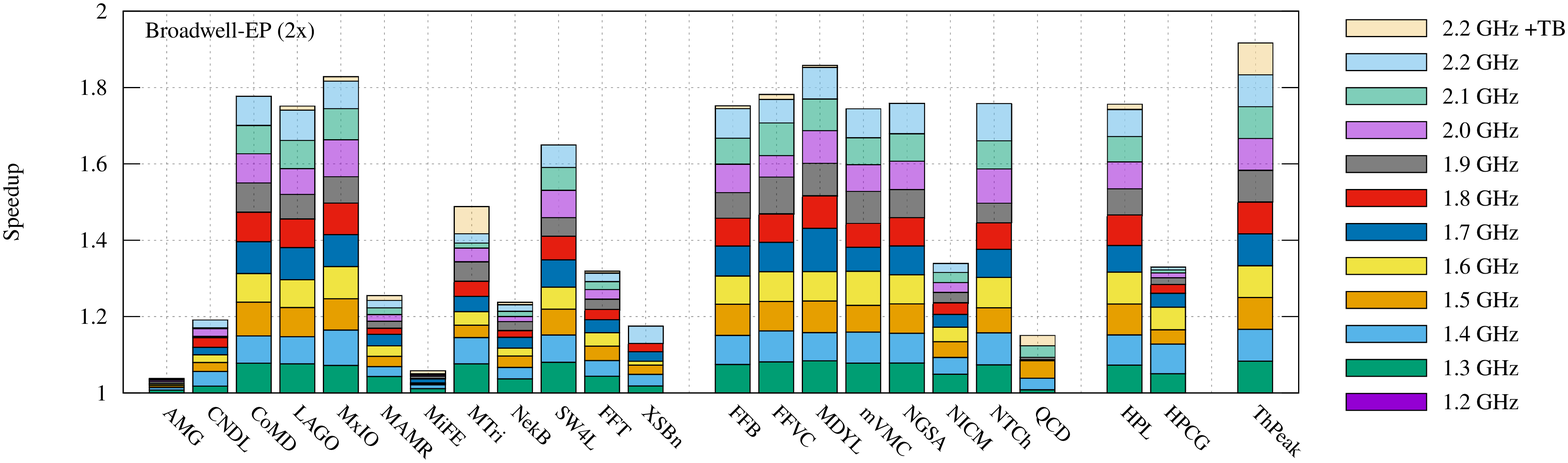}
    \vspace*{-5mm}
    \caption{\label{fig:freq} Speedup obtained through increased CPU frequency (w.r.t baseline frequency of \unit[1.0]{GHz} on KNL/KNM and \unit[1.2]{GHz} on BDW); \textbf{Top plot: KNL, middle plot: KNL, bottom plot: BDW}; Theoretical peak (ThPeak): furthest right bar; Labels/abbreviations of proxy-apps according to Section~\ref{ssec:bm} and 'TB' = Turbo Boost is assumed to be \unit[+100]{Mhz} across all cores}
    \vspace{-0.2em}
\end{figure}

For this test, we disable turbo boost and throttle the core frequency, but keep the uncore at maximum frequency which would otherwise negatively
affect the memory subsystem, to identify each application's dependency on ALU/FPU performance. The shown speedup (w.r.t time-to-solution) in Figure~\ref{fig:freq}
for each proxy-app is relative to the lowest CPU frequency on each architecture, and we include our \textit{performance} results
(cf. Section~\ref{ssec:bmconf}) with maximum frequency plus enabled turbo boost (labeled with ``+TB'').
\added{
It should be noted, that Intel abandoned single-frequency turbo boost long ago, and the real TB frequency band
depends on multiple factors~\cite{lento_whitepaper:_2014}, such as \#cores, utilized units, etc. Hence, we choose an universal, but
pessimistic \unit[+100]{Mhz} (cf.~Table~\ref{table:HW}) for the TB plot in this figure, and therefore
an application may exceed our pessimistic peak, as it is evident for Knights Landing in the top plot.
}

While a benefit from enabled turbo boost on BDW is near invisible (except for MTri), the proxy-apps clearly reduce their
time-to-solution on KNL and KNM when these CPUs are allowed to turbo. Overall, the benchmarks seem to be less memory-bound
and more compute-bound, especially salient for AMG and MiniFE,  when moving to Xeon~Phi, indicating a clear benefit
from the much bigger/faster MCDRAM used as last-level cache and indicating a more balanced
(w.r.t bandwidth to \unit[]{flop/s} ratio) architecture.
However, the limited speedup for HPL on KNL clearly shows the CPU's abundance of FP64 units.
Here, the successor, Knights Mill, shows a better balance.
Another interesting observation is the inverse behavior of AMG and HPCG on our tested architecture.
Both benchmarks are supposed to be memory-bound, but the absence of signs of any scalability with frequency on Xeon~Phi
strengthens our hypothesis from Section~\ref{ssec:eval_mem} that HPCG is primarily bound by memory latency.
\begin{figure}[tbp]
    \centering
    \includegraphics[width=\linewidth]{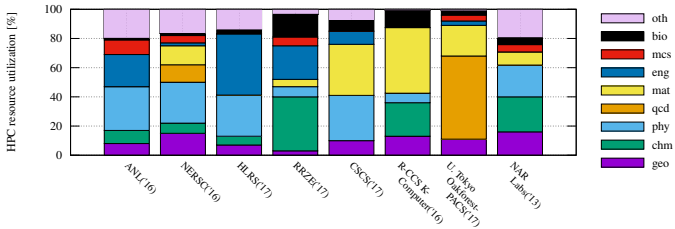}
    \vspace*{-7mm}
    \caption{\label{fid:disc:breakdown} Annual HPC site/system utilization by domain; Labels acc. to Table~\ref{table:APP}: \texttt{geo} = Geo-/Earthscience, \texttt{chm} = Chemistry, \texttt{phy} = Physics, \texttt{qcd} = Lattice QCD, \texttt{mat} = Material Science/Engineering, \texttt{eng} = Engineering (Mechanics, CFD), \texttt{mcs} = Math/Computer Science, \texttt{bio} = Bioscience, \texttt{oth} = \textit{Other}}
    \vspace{-1em}
\end{figure}
%

For I/O portions of an application, the Figure~\ref{fig:freq} reveals another observation, i.e., MACSio's write speed
scales with increased frequency. Since, MACSio performs only single figure \unit[]{GIop/s} and negligible
\unit[]{flop/s}, increasing the CPU's compute capabilities cannot explain the shown speedup.
Hence, our theory is: MACSio (and I/O in general) is bound by the Linux kernel, whose performance depends on CPU frequency.
Gu{\'e}rout et al. report similar findings~\cite{guerout_energy-aware_2013}, and we see equivalent behavior with 
a micro-benchmark (with Unix's \texttt{dd} command).

\subsection{Remaining Metrics}\label{ssec:eval_rest}
To disseminate the remaining results from our experiments, we 
attached Table~\ref{table:rest} to this paper, which can be utilized for
further analysis, and which contains some interesting data points.
For example, the power measurements for CANDLE, the results of which are just slightly
higher in comparison to MACSio, indicate that Intel's MKL-DNN (used
underneath to compute on the FP16 VNNI units for KNM or FP32 units for KNL) does not
fully utilize the CPUs' potential. Furthermore, the L2 hit rate on both
Xeon Phi systems is considerably higher than on our reference
hardware, indicating improvements in the hardware prefetcher and are presumably
a direct effect of the high-bandwidth MCDRAM which is used in \textit{cache} mode.

\section{Discussion and Implications}\label{sec:discuss}
%
%

While the previous section focuses on the collected data and comparisons between
the three architectures, this section summarizes the relevant points to consider
from our study, which should be taken into account when moving forward.

\subsection{Performance Metrics}
The de facto performance metric reported in HPC is \unit[]{flop/s}. However, reporting \unit[]{flop/s} is not limited to applications that are compute-bound. Benchmarks that are designed to resemble realistic workloads, e.g., the
memory-bound HPCG benchmark, typically report performance in \unit[]{flop/s}. The proxy-/mini-apps in this study
as well typically report \unit[]{flop/s} despite the fact that only six out of 20 proxy-/mini-apps we analyze in
this study appear to be compute-bound (including NGSA that is bound by ALUs, not FPUs). We
argue that convening on reporting relevant metrics would shift the focus of the community to be less \unit[]{flop/s}-centered.

\subsection{Considerations for HPC Utilization by Scientific Domain}\label{ssec:workload_util}
%
%
This paper highlights the diminishing relevance of \unit[]{flop/s} when
considering the actual requirements of representative proxy-apps.
The relevance of \unit[]{flop/s} on a given supercomputer can be further
diminished when considering the analysis of node-hours spent yearly on
different scientific domains at supercomputing facilities.
Figure~\ref{fid:disc:breakdown} summarizes the breakdown of node-hours by
scientific domain for different supercomputing facilities (based on yearly
reports of mentioned facilities). For instance, by simply mapping the scientific
domains in Figure~\ref{fid:disc:breakdown} to representative proxies,
ANL's ALCF and \mbox{R-CCS's} K-computer would be achieving $\approx$14\% and
$\approx$11\% of the peak \unit[]{flop/s}, respectively, when projecting
for the annual node-hours. 
It is worth mentioning that the relevance of \unit[]{flop/s} is even more
of an issue for supercomputers dedicated to specific workloads: the relevance of
\unit[]{flop/s} can vary widely. For instance, a supercomputer dedicated
mainly to weather forecasting, e.g., the~\unit[18]{Pflop/s} system recently
installed at Japan's Meteorological Agency~\cite{japan_meteorological_agency_jma_jma_2018},
should give minimal relevance to \unit[]{flop/s} since the proxy representing
this workload on that supercomputer achieves $\approx$6\% of the peak \unit[]{flop/s},
because those workloads are typically memory-bound. On the other hand, a
supercomputer dedicated to AI/ML such as ABCI, the world's 5\textsuperscript{th}
fastest supercomputer as of June 2018, would put high emphasis on \unit[]{flop/s}
due to the fact that current deep learning workloads rely heavily on dense matrix multiplications.

\subsection{Memory-bound Applications}
As demonstrated in Figure~\ref{fig:flops}, the performance of memory-bound
applications is mostly not affected by the peak \unit[]{flop/s} available.
Accordingly, investment in data-centric architectures and programming models
should take priority over paying premium for \unit[]{flop/s}-centric systems.
In one motivating instance, an investigation conducted by the NASA Ames Research Center,
for a planned upgrade of the Pleiades supercomputer in 2016~\cite{saini_performance_2016},
concluded that the performance gain of their applications from upgrading to
Intel Haswell processors was insignificant in comparison to using the older
Ivy Bridge-based processors (the newer processor offered double the peak
\unit[]{flop/s} at almost the same memory bandwidth).
And hence the choice was to only do a partial upgrade to Haswell processors.

\subsection{Compute-bound Applications}
Investing more in data-centric architectures to accommodate memory-bound
applications can have a negative impact on the remaining minority of
applications: compute-bound applications. Considering the market trends that
are already pushing away from dedicating the majority of chip area to
FP64 units, it is likely that libraries with compute-bound code (e.g., BLAS)
would support mixed precision or emulation by lower precision FPUs. The
remaining applications that do not rely on external libraries might suffer a
performance hit.




\section{Related Work}\label{sec:relwork}
%
Apart from RIKEN's mini-apps and the ECP proxy-apps, which we use for our study, there are numerous benchmark suites
based on proxy applications from other HPC centers and institutes available
~\cite{prace_unified_2016,noauthor_mantevo_nodate,nersc_characterization_nodate,llnl_llnl_nodate,llnl_coral_nodate,spec_spec_nodate}.
Overall those lists show a partial overlap, either directly (i.e., same benchmark) or indirectly (same
scientific domain), between all these suites, which, for example, were used to analyze message passing characteristic~\cite{klenk_overview_2017} or to assess how predictable full application performance is
based on proxy-app measurements~\cite{barrett_assessing_2015}.
Hence, our systematic approach and published framework \textbf{\url{https://gitlab.com/domke/PAstudy}} can be
transferred to these alternative benchmarks for complementary studies, and our included raw data
can be investigated further w.r.t metrics which were outside the scope of our study.

Furthermore, the HPC community has already started to analyze relevant workloads with respect to arithmetic
intensity or memory and other potential bottlenecks for some proxy-apps~
\cite{aaziz_methodology_2018,asifuzzaman_report_2017,koskela_novel_2018}
and individual applications~\cite{culpo_current_2012,tramm_memory_2015,kumahata_kernel_2013},
revealing similar results to ours that most realistic HPC codes are not compute-bound and
achieve very low computational efficiency,
which in demonstrated cases affected procurement decisions~\cite{saini_performance_2016}.
However, to the best of our knowledge, we are the first to present a broad study across a
wide spectrum of HPC workloads which aims at characterizing bottlenecks and aims
specifically at identifying floating-point unit/precision requirements for modern architectures.


\section{Conclusion}\label{sec:conclusion}
%

We compared two architectural similar processors that have different double-precision
silicon budget. By studying a large number of HPC proxy application, we found no significant 
performance difference between these two processors, despite one having more double-precision
compute than the other. Our study points toward a growing need to re-iterate and re-think
architecture design decisions in high-performance computing, especially with respect
to precision. 
Do we really need the amount of double-precision compute that modern processors offer?
Our results on the Intel Xeon Phi twins points towards a 'No', and we hope that this
work inspires other researchers to also challenge the floating-point to silicon distribution
for the available and future general-purpose processors, graphical processors, or
accelerators in HPC systems.


\iftoggle{includeacknowl}{
  \section*{Acknowledgment \& Author Contributions}
\added{
This work was supported by MEXT, JST special appointed survey 30593/2018, JST-CREST Grant Number JPMJCR1303,
JSPS KAKENHI Grant Number JP16F16764, the New Energy and Industrial Technology Development
Organization (NEDO), and the AIST/TokyoTech Real-world Big-Data Computation Open Innovation Laboratory (RWBC-OIL).
Moreover, we thank Intel for their technical support.
The authors~K.M., J.D., H.Z., K.Y., T.T. and Y.T. performed the experiments and data collection.
J.D., M.W., A.P. designed the study, analyzed the data, and supervised its execution together
with S.M., while all authors contributed to writing and editing.
}

}{
}

\bibliographystyle{IEEEtran}
\bibliography{IEEEabrv,precision}
%
%
%


\iftoggle{includeappendix}{
    \clearpage
    \appendices
%
%
%
%
%
\section{Additionally Evaluated Metrics}\label{apx:metrics}
%
%
\begin{table*}[tbp]
    \caption{\label{table:rest} Application configuration and measured metrics; Missing data for CANDLE due to SDE crashes on Phi; Measurements indicate CANDLE/MKL-DNN ignores OpenMP settings and tries to utilize full chip $\rightarrow$ listed in italic; Label explanation: t2sol = time-to-solution (kernel), Gop (D $|$ S $|$ I) = Giga operations (FP64 $|$ FP32 $|$ Integer), SIMDi/cyc = SIMD instructions per cycle, FPAIp[R $|$ W] = FP Arithmetic instructions per memory [read $|$ write], [B $|$ M]Bd = [Back-end $|$ Memory] Bound (see~\cite{sobhee_intel_2018} for details), L2h = L2 cache hit rate, LLh = Last level cache hit rate (L3 for BDW, MCDRAM for KNL/KNM), Gbra/s = Giga branches/s;\qquad\added{Note: SIMDi/cyc and FPAIp* as well as BBd and MBd occupy the same columns due to their similarity and space constraints}}
    \centering\scriptsize
    \begin{tabular}{|l|r|r|r|r|r|r|r|c|r|r|r|r|}
        \hline \hC
        \tH{\textbf{KNL}} & \tH{\#MPI} & \tH{\#OMP} & \tH{t2sol [\unit[]{s}]} & \tH{\#Gop (D)} & \tH{\#Gop (S)} & \tH{\#Gop (I)} &  \tH{Power [\unit[]{W}]} & \tH{\#SIMDi/cyc} & \tH{BBd [\%]} & \tH{L2h [\%]} & \tH{LLh [\%]} & \tH{Gbra/s} \\ \hline
        AMG	        &	1	&	128	&	6.057	&	110.271	&	0	&	352.640	&	202.16	&	0.063	&	77.3	&	93	&    74.6	&	7.310	\\ \hline \rC
        CANDLE	    &	1	&	\textit{32}	&	59.796	&	\textit{N/A}	&	\textit{N/A}	&	\textit{N/A}	&	143.79	&	0.105	&	67.4	&	86	&    89.7	&	\textit{N/A}	\\ \hline
        CoMD	    &	32	&	8	&	3.199	&	161.691	&	14.842	&	3.476	&	189.24	&	0.077	&	81.0	&	85	&    99.4	&	11.551	\\ \hline \rC
        Laghos	    &	64	&	4	&	13.508	&	85.547	&	0.422	&	1055.977	&	143.84	&	0.021	&	23.7	&	98	&    99.7	&	7.839	\\ \hline
        MACSio	    &	64	&	1	&	35.110	&	0.613	&	0.007	&	77.884	&	140.02	&	0.002	&	52.8	&	98	&    98.6	&	18.115	\\ \hline \rC
        miniAMR	    &	128	&	1	&	47.150	&	291.536	&	0.014	&	3358.569	&	153.44	&	0.009	&	75.9	&	71	&    97.6	&	5.023	\\ \hline
        miniFE	    &	1	&	256	&	0.694	&	28.961	&	0	&	177.704	&	221.69	&	0.022	&	81.8	&	93	&    93.6	&	10.930	\\ \hline \rC
        miniTri	    &	1	&	128	&	8.630	&	0	&	0	&	118.261	&	131.49	&	0.001	&	81.9	&	66	&    99.5	&	4.531	\\ \hline
        Nekbone	    &	128	&	1	&	3.290	&	410.361	&	0	&	23.371	&	221.48	&	0.050	&	76.5	&	87	&    97.6	&	6.125	\\ \hline \rC
        SW4lite	    &	64	&	4	&	1.686	&	145.938	&	0	&	0.761	&	214.57	&	0.096	&	80.6	&	95	&    98.4	&	2.218	\\ \hline
        SWFFT	    &	128	&	1	&	1.235	&	12.688	&	0.005	&	42.509	&	174.12	&	0.029	&	76.7	&	83	&    98.6	&	20.905	\\ \hline \rC
        XSBench	    &	1	&	256	&	1.290	&	27.283	&	0.417	&	16.441	&	192.46	&	0.041	&	93.7	&	22	&    99.5	&	2.629	\\ \hline\hline
        FFB	        &	64	&	2	&	8.244	&	2.300	&	258.561	&	1785.716	&	179.55	&	0.159	&	38.4	&	89	&    99.7	&	2.717	\\ \hline \rC
        FFVC	    &	1	&	64	&	13.009	&	134.589	&	1579.917	&	20174.483	&	180.58	&	0.169	&	36.0	&	95	&    99.7	&	4.415	\\ \hline
        mVMC	    &	32	&	6	&	20.679	&	1141.865	&	1.345	&	1746.001	&	180.98	&	0.036	&	81.9	&	91	&    98.9	&	6.073	\\ \hline \rC
        MODYLAS	    &	64	&	4	&	22.514	&	6287.279	&	2.063	&	23104.728	&	206.98	&	0.072	&	80.4	&	97	&    95.7	&	7.742	\\ \hline
        NGSA	    &	4	&	32	&	829.675	&	0.826	&	0.023	&	69.117	&	97.91	&	0.002	&	51.9	&	71	&    95.9	&	1.050	\\ \hline \rC
        NICAM	    &	10	&	15	&	37.802	&	422.504	&	0.066	&	925.228	&	119.46	&	0.193	&	67.8	&	92	&    99.2	&	0.231	\\ \hline
        NTChem	    &	16	&	8	&	18.985	&	1629.210	&	0.627	&	2303.804	&	167.13	&	0.060	&	64.4	&	91	&    99.2	&	5.429	\\ \hline \rC
        QCD	        &	1	&	128	&	8.437	&	631.522	&	0	&	3823.335	&	215.67	&	0.220	&	69.4	&	88	&    95.4	&	1.151	\\ \hline\hline
        HPCG	    &	96	&	1	&	44.612	&	612.799	&	0	&	17530.136	&	181.69	&	0.023	&	86.1	&	91	&    45.7	&	1.446	\\ \hline \rC
        HPL	        &	64	&	1	&	145.400	&	184191.774	&	0.015	&	20226.567	&	221.13	&	0.374	&	52.3	&	93	&    87.9	&	1.232	\\ \hline
        \hline \hC
        \tH{\textbf{KNM}} & \tH{\#MPI} & \tH{\#OMP} & \tH{t2sol [\unit[]{s}]} & \tH{\#Gop (D)} & \tH{\#Gop (S)} & \tH{\#Gop (I)} &  \tH{Power [\unit[]{W}]} & \tH{\#SIMDi/cyc} & \tH{BBd [\%]} & \tH{L2h [\%]} & \tH{LLh [\%]} & \tH{Gbra/s} \\ \hline
        AMG	        &	1	&	128	&	7.434	&	110.271	&	0	&	352.639	&	202.52	&	0.062	&	75.4	&	94	&   73.3	&	6.392	\\ \hline \rC
        CANDLE	    &	1	&	\textit{144}	&	50.527	&	\textit{N/A}	&	\textit{N/A}	&	\textit{N/A}	&	153.69	&	0.040	&	82.4	&	92	&	90.9	&	\textit{N/A}	\\ \hline
        CoMD	    &	72	&	2	&	3.194	&	161.842	&	14.880	&	3.479	&	196.64	&	0.177	&	67.5	&	86	&	99.1	&	11.546	\\ \hline \rC
        Laghos	    &	64	&	4	&	12.725	&	85.383	&	0.422	&	1056.141	&	139.33	&	0.023	&	25.1	&	98	&	99.8	&	8.345	\\ \hline
        MACSio	    &	64	&	1	&	33.236	&	0.613	&	0.007	&	77.884	&	135.48	&	0.002	&	53.8	&	98	&	98.2	&	19.206	\\ \hline \rC
        miniAMR	    &	128	&	1	&	44.653	&	291.536	&	0.014	&	3358.570	&	177.31	&	0.009	&	75.3	&	71	&	97.3	&	5.337	\\ \hline
        miniFE	    &	72	&	1	&	0.669	&	32.892	&	0	&	669.371	&	210.18	&	0.097	&	55.6	&	60	&	98.3	&	7.393	\\ \hline \rC
        miniTri	    &	1	&	128	&	9.545	&	0	&	0	&	118.262	&	122.02	&	0	&	80.9	&	68	&	99.6	&	4.102	\\ \hline
        Nekbone	    &	144	&	1	&	2.984	&	410.381	&	0	&	23.470	&	233.46	&	0.040	&	76.1	&	87	&	96.6	&	6.494	\\ \hline \rC
        SW4lite	    &	72	&	4	&	1.569	&	146.048	&	0	&	0.764	&	228.01	&	0.090	&	81.3	&	96	&	97.8	&	2.753	\\ \hline
        SWFFT	    &	128	&	1	&	1.189	&	12.555	&	0.005	&	41.732	&	172.66	&	0.026	&	77.2	&	83	&	98.5	&	21.990	\\ \hline \rC
        XSBench	    &	1	&	288	&	1.220	&	30.603	&	0.417	&	16.440	&	197.16	&	0.038	&	91.5	&	22	&	98.5	&	2.783	\\ \hline\hline
        FFB	        &	64	&	2	&	7.750	&	2.300	&	258.565	&	1785.712	&	178.72	&	0.171	&	38.6	&	89	&	99.7	&	2.886	\\ \hline \rC
        FFVC	    &	1	&	72	&	13.497	&	134.589	&	1579.917	&	20174.587	&	182.05	&	0.162	&	55.2	&	94	&	99.9	&	5.055	\\ \hline
        mVMC	    &	72	&	4	&	19.659	&	1140.670	&	1.347	&	1802.663	&	197.64	&	0.012	&	76.0	&	91	&	98.5	&	8.869	\\ \hline \rC
        MODYLAS	    &	64	&	4	&	24.026	&	6287.279	&	2.063	&	23104.728	&	217.47	&	0.062	&	80.0	&	97	&	95.6	&	7.153	\\ \hline
        NGSA	    &	4	&	18	&	724.546	&	0.826	&	0.023	&	69.300	&	88.67	&	0.002	&	39.5	&	68	&	94.9	&	1.138	\\ \hline \rC
        NICAM	    &	10	&	7	&	34.380	&	422.504	&	0.066	&	925.229	&	113.88	&	0.208	&	68.2	&	92	&	99.1	&	0.248	\\ \hline
        NTChem	    &	72	&	2	&	14.606	&	1575.310	&	0.623	&	1985.255	&	176.51	&	0.066	&	59.0	&	90	&	98.4	&	7.038	\\ \hline \rC
        QCD	        &	1	&	144	&	9.662	&	631.522	&	0	&	3823.337	&	200.86	&	0.175	&	72.6	&	88	&	95.9	&	2.121	\\ \hline\hline
        HPCG	    &	64	&	1	&	42.865	&	612.605	&	0	&	17532.326	&	174.58	&	0.041	&	86.5	&	95	&   42.9	&	2.878	\\ \hline \rC
        HPL	        &	72	&	1	&	146.562	&	184893.073	&	0.016	&	20414.548	&	263.59	&	0.351	&	57.0	&	92	&   87.0	&	1.885	\\ \hline
        \hline \hC
        \tH{\textbf{BDW}} & \tH{\#MPI} & \tH{\#OMP} & \tH{t2sol [\unit[]{s}]} & \tH{\#Gop (D)} & \tH{\#Gop (S)} & \tH{\#Gop (I)} & \tH{Power [\unit[]{W}]} & \tH{FPAIp[R : W]} & \textcolor{white}{MBd [\%]} & \tH{L2h [\%]} & \tH{LLh [\%]} & \tH{Gbra/s} \\ \hline
        AMG	        &	8	&	6	&	10.780	&	110.810	&	0	&	362.209	&	152.21	&	0.361 : 5.516	&	44.8	&	21		&	17  &	4.354	\\ \hline \rC
        CANDLE	    &	1	&	\textit{12}	&	78.240	&	0.012	&	6918.340	&	2783.532	&	132.38	&	1.078 : 2.800	&	26.7	&	23		&    11	&	1.242	\\ \hline
        CoMD	    &	48	&	1	&	2.921	&	152.022	&	0	&	0.205	&	133.17	&	0.845 : 6.615	&	1.5		&	15		&    15	&	11.391	\\ \hline \rC
        Laghos	    &	24	&	1	&	5.5472	&	44.534	&	0	&	421.465	&	126.51	&	0.184 : 0.476	&	13.2	&	81		&    56	&	16.808	\\ \hline
        MACSio	    &	4	&	1	&	10.498	&	0.070	&	0	&	72.582	&	89.3	&	0 : 0	&	0.8		&	48		&    59	&	3.274	\\ \hline \rC
        miniAMR	    &	96	&	1	&	55.386	&	40.816	&	0	&	172.317	&	133.29	&	0.059 : 0.311	&	55.1	&	24		&    23	&	4.013	\\ \hline
        miniFE	    &	24	&	1	&	1.475	&	30.693	&	0	&	120.715	&	152.77	&	0.311 : 5.454	&	55.2	&	15		&    12	&	4.699	\\ \hline \rC
        miniTri	    &	1	&	48	&	5.478	&	0	&	0	&	118.178	&	112.61	&	0 : 0	&	34.0	&	47		&	90    &	7.106	\\ \hline
        Nekbone	    &	96	&	1	&	5.671	&	301.559	&	0	&	10.139	&	154.74	&	0.593 : 2.431	&	36.9	&	36		&	24    &	3.915	\\ \hline \rC
        SW4lite	    &	24	&	2	&	2.056	&	136.835	&	0	&	1.585	&	146.65	&	1.044 : 4.580	&	9.1		&	75		&    18	&	1.112	\\ \hline
        SWFFT	    &	32	&	1	&	1.088	&	12.239	&	0	&	38.782	&	134.55	&	0.117 : 0.675	&	28.3	&	23		&    32	&	20.932	\\ \hline \rC
        XSBench	    &	1	&	96	&	2.022	&	19.921	&	0	&	20.280	&	132.25	&	0.807 : 3.847	&	71.7	&	5		&    18	&	1.653	\\ \hline\hline
        FFB	        &	24	&	1	&	5.327	&	1.300	&	233.640	&	2116.421	&	144.35	&	0.635 : 2.200	&	21.3	&	79		&    33	&	3.723	\\ \hline \rC
        FFVC	    &	12	&	4	&	12.691	&	127.322	&	1573.782	&	27857.376	&	151.85	&	0.481 : 2.844	&	3.3		&	84		&    57	&	9.045	\\ \hline
        mVMC	    &	24	&	2	&	13.489	&	1092.394	&	0	&	2224.092	&	152.28	&	0.601 : 2.456	&	12.0	&	36		&    24	&	10.170	\\ \hline \rC
        MODYLAS	    &	16	&	3	&	36.101	&	5363.366	&	0	&	10888.745	&	135.75	&	0.875 : 8.736	&	8.1		&	60		&    31	&	5.385	\\ \hline
        NGSA	    &	12	&	4	&	105.879	&	0.826	&	0.023	&	64.249	&	107.15	&	0.002 : 0.006	&	6.5		&	21		&    36	&	8.566	\\ \hline \rC
        NICAM	    &	10	&	6	&	28.449	&	428.282	&	0.003	&	687.852	&	118.32	&	0.540 : 3.732	&	49.6	&	27		&    19	&	0.585	\\ \hline
        NTChem	    &	24	&	1	&	8.963	&	1315.509	&	0	&	778.829	&	141.3	&	0.867 : 4.931	&	9.4		&	56		&    39	&	10.173	\\ \hline \rC
        QCD	        &	1	&	24	&	13.102	&	612.303	&	0	&	3817.944	&	153.2	&	1.152 : 4.542	&	45.2	&	27		&    24	&	0.368	\\ \hline\hline
        HPCG	    &	2	&	24	&	38.595	&	559.046	&	0	&	90.171	&	166.18	&	0.143 : 0.628	&	11.3	&	34		&    23		&	10.928	\\ \hline \rC
        HPL	        &	24	&	1	&	271.794	&	181484.240	&	0	&	31919.479	&	189.37	&	\,~~2.280 : 122.693	&	3.9		&	10		&    3    	&	2.147	\\ \hline
    \end{tabular}
\end{table*}

}{
    
}

\end{document}